\documentclass{article}

\usepackage{hyperref}                 
\hypersetup{
	hidelinks,                        
	colorlinks = true,
	citecolor = cyan,
	urlcolor = black,
	linkcolor = magenta
}
\usepackage{arxiv}
\usepackage{amsmath}
\usepackage[utf8]{inputenc} 
\usepackage[T1]{fontenc}    
\usepackage{url}            
\usepackage[numbers]{natbib}  
\usepackage{doi}             
\usepackage{booktabs}       
\usepackage{amsfonts}       
\usepackage{amssymb}        
\usepackage{nicefrac}       
\usepackage{microtype}      
\usepackage{cleveref}       
\usepackage{graphicx}
\graphicspath{{./}}
\usepackage{natbib}
\usepackage{doi}
\usepackage{xcolor}
\usepackage{enumerate}

\usepackage{subcaption}
\usepackage{threeparttable} 

\title{Reduced-Precision Stochastic Simulation for Mathematical Biology}

\renewcommand{\shorttitle}{Reduced-Precision Stochastic Simulation}

\usepackage{authblk}

\newbox{\orcid}\sbox{\orcid}{\includegraphics[scale=0.06]{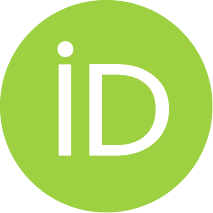}}

\author[1,2]{%
	\href{https://orcid.org/0000-0002-6542-6032}{\usebox{\orcid}\hspace{1mm}Tom Kimpson\thanks{Corresponding author: \texttt{tom.kimpson@unimelb.edu.au}}}%
}

\author[3]{%
	\href{https://orcid.org/0000-0002-4697-4789}{\usebox{\orcid}\hspace{1mm}Mark B. Flegg}%
}

\author[1,2]{%
	\href{https://orcid.org/0000-0002-8809-726X}{\usebox{\orcid}\hspace{1mm}Jennifer A. Flegg}%
}

\affil[1]{School of Mathematics and Statistics, The University of Melbourne, Parkville, VIC 3010, Australia}
\affil[2]{ARC Centre of Excellence for the Mathematical Analysis of Cellular Systems (MACSYS), The University of Melbourne, Parkville, VIC 3010, Australia }

\affil[3]{School of Mathematics, Monash University, Clayton, VIC 3800, Australia}

\renewcommand{\headeright}{}

\begin{document}
\maketitle

\begin{abstract}
The stochastic simulation algorithm (SSA) is widely used to perform exact forward simulation of discrete stochastic processes in biology. However, the computational cost, driven by sequential event-by-event sampling across large ensembles, remains a computational barrier. We investigate whether reduced-precision floating-point arithmetic can accelerate SSA without degrading statistical fidelity, drawing on the success of reduced-precision methods in weather and climate modelling. We evaluate two strategies across five canonical models (birth--death, Schl\"{o}gl, Telegraph, dimerisation, repressilator): (i) mixed precision, computing propensities in 16-bit while maintaining accumulators in 32-bit; and (ii) uniform precision, performing all arithmetic in 16-bit. Mixed-precision SSA produces ensemble statistics that closely match the 64-bit reference for all models, as measured by Kolmogorov--Smirnov tests and Wasserstein distances. Under uniform precision, deterministic rounding introduces systematic biases across several models, with catastrophic failures in some cases. Stochastic rounding (SR) and propensity normalisation eliminate these biases, restoring distributional fidelity across all models tested (KS $p > 0.05$). Our results establish mixed-precision SSA with SR as a viable acceleration strategy for mathematical biology: 16-bit formats shrink per-variable data size by $2$--$4\times$ relative to \texttt{fp32}/\texttt{fp64}, yielding comparable reductions in memory footprint and up to $\sim 1.5\times$ wall-clock speedup on CPU hardware that lacks native 16-bit arithmetic. As a hardware-level acceleration, mixed-precision SSA complements algorithmic methods such as tau-leaping and maps naturally onto modern GPU and TPU architectures with native 16-bit arithmetic.
\end{abstract}

\keywords{Stochastic simulation algorithm \and Reduced precision \and Stochastic rounding \and Mixed precision \and Mathematical biology}

\section{Introduction}
Stochastic simulation is a central tool in computational systems biology. From minimal gene circuits to whole-cell models, the stochastic simulation algorithm (SSA) and its variants provide exact forward simulation of discrete stochastic processes, generating sample paths of the chemical master equation one reaction event at a time. Yet the computational cost remains substantial: realistic models may involve hundreds of species and reactions, require long trajectories, large ensembles, and repeated parameter sweeps or inference loops.
Traditional acceleration strategies have focused on algorithmic modifications such as tau-leaping, R-leaping, hybrid multiscale solvers, and quasi–steady-state reductions \cite{gillespie1977,anderson2007,cao2006,hybrid1,hybrid2}. These methods often deliver order-of-magnitude speedups, but at the price of additional modelling choices and potential approximation bias, especially for stiff networks or rare events.

This paper targets a different axis for computational speed-up: reduced-precision arithmetic. Hardware trends driven by machine learning have made 16-bit floating-point formats (IEEE \texttt{fp16} and \texttt{bfloat16}) ubiquitous on modern GPUs and TPUs \cite{jouppi2017,kalamkar2019}. By shrinking the data size of each number by $2-4\times$ relative to \texttt{fp32}/\texttt{fp64}, these formats cut memory traffic by the same factor, allowing more data to fit in fast on-chip memory and reducing energy consumption. The question is whether SSA can exploit this without sacrificing statistical fidelity.

We argue the answer is yes, by analogy to numerical weather and climate modelling, where reduced precision is now well established. In those domains, reduced precision is used because rounding error is typically dominated by observation noise, subgrid parameterisations, and chaotic dynamics \cite{palmer2015,chantry2019,dawson2018}. Mixed-precision schemes (compute in 16-bit, accumulate/update in 32-bit) accelerate throughput without degrading ensemble statistics, provided numerically sensitive operations (e.g., conserved quantities, long reductions, large–small cancellations) are protected. Stochastic rounding (SR) \cite{parker1997,connolly2021,croci2022,croci2022review,kimpson2024} further mitigates precision loss by replacing deterministic round-to-nearest with unbiased probabilistic rounding, preventing systematic drift when small values are repeatedly added to large accumulators. In climate models, the combination of mixed precision and SR has proven effective for maintaining long-term statistical accuracy \cite{thornes2017,paxton2022}.

Stochastic biological simulation sits in an analogous regime. Intrinsic noise from finite copy numbers \cite{elowitz2002}, parametric uncertainty from poorly constrained rates \cite{gutenkunst2007}, and structural uncertainty from incomplete networks \cite{babtie2014} may exceed any bias from low-precision arithmetic. The numerically hazardous cases (long reductions, invariants, large–small updates) are localised and predictable \cite{connolly2021}. Moreover, SSA is often memory-bound rather than compute-bound: each step performs modest arithmetic while moving data irregularly across species counts, rate constants, and stoichiometry structures \cite{cao2004}. In complexity terms, a typical step executes $\mathcal{O}(1)$ transcendental/log operations but streams sparsely over $\mathcal{O}(N+M)$ state and propensity structures (where $N$ is the number of species and $M$ is the number of reactions) making memory bandwidth the throughput bottleneck rather than floating-point operations \cite{gibson2000,slepoy2008}. Shrinking data representations directly reduces the dominant cost, making mixed precision a natural target for SSA acceleration. This is especially relevant for large-scale applications such as whole-cell models \cite{karr2012}, where thousands of species and reactions amplify the memory-bandwidth bottleneck.

This opportunity remains under-explored in mathematical biology. While reduced precision has gained traction in weather and climate modelling, no systematic study has examined whether SSA can achieve similar gains in mathematical biology. The key questions are: (i) does the statistical fidelity of ensemble simulations degrade under reduced precision? (ii) which operations require protection, and does SR suffice to prevent systematic bias? and (iii) how much speedup and memory reduction can 16-bit formats deliver for large-scale stochastic models?

We evaluate mixed-precision SSA on canonical model systems chosen to stress different numerical regimes: a simple birth-death process to establish a baseline; the Schlogl model for its bistability and rare-event dynamics; the Telegraph process to probe bursty, multi-timescale behaviour; a dimerisation network for its nonlinear mass-action kinetics; and the repressilator for its sustained, stiff oscillations. Table~\ref{tab:taxonomy} situates our approach relative to existing SSA acceleration strategies. These test cases allow us to assess whether precision degradation depends on model structure, and whether safe "precision maps" (rules for assigning precision to variables and operations based on numerical sensitivity) can be identified.

Although we focus on the SSA as the simplest and most widely used stochastic framework, the core question, whether intrinsic stochastic noise tolerates reduced-precision arithmetic, applies to other simulation paradigms in mathematical biology. Spatial stochastic methods such as the reaction-diffusion master equation (RDME) \cite{engblom2009}, which couples SSA-like reactions within lattice subvolumes to stochastic diffusive hopping between neighbours, face similar memory-bandwidth bottlenecks amplified by spatial degrees of freedom. Agent-based models, widely used in epidemiology and tissue modelling \cite{an2009}, share the same discrete stochastic character. Even deterministic reaction-diffusion PDE solvers \cite{hundsdorfer2003}, though lacking intrinsic noise, are memory-bound on fine grids and could benefit from halved data traffic. The well-stirred SSA is the natural starting point: it is both the simplest stochastic framework and the kernel on which RDME and hybrid spatial methods are built, so establishing reduced-precision fidelity here is a prerequisite for extending the approach to spatially resolved settings, which we leave to future work.

We ground the investigation in a Julia implementation for clarity and reproducibility. Julia's type system allows us to cleanly separate precision choices: we use 16-bit types (\texttt{bf16}/\texttt{fp16}) for propensities and intermediate computations, while keeping accumulators (summations, time updates) in \texttt{Float32}. When needed, we introduce lightweight SR-wrapped types that override basic arithmetic with SR semantics. This design makes ablations straightforward: we can swap \texttt{fp16} for \texttt{bf16}, toggle SR on or off, and vary accumulator precision without changing control flow.

The remainder of this paper is organised as follows. \Cref{sec:ssa} introduces the SSA and identifies the operations most sensitive to floating-point precision. \Cref{sec:rp_sr} reviews reduced-precision floating-point formats and SR, and describes our strategy for assigning precision to different SSA operations. \Cref{sec:experiments} presents numerical experiments on five canonical biochemical kinetics models of increasing complexity, comparing six precision configurations against a double-precision reference. \Cref{sec:discussion} discusses the implications of these results, including practical guidelines for practitioners and limitations of the current study. \Cref{sec:conclusion} concludes with a summary of our findings and directions for future work.

\begin{table}[t]
	\centering
	\small
	\begin{tabular}{@{}p{2.8cm}p{3.2cm}p{3.5cm}p{4.5cm}@{}}
		\toprule
		\textbf{Strategy} & \textbf{Core Principle} & \textbf{Primary Advantage} & \textbf{Key Limitation / Trade-off} \\
		\midrule
		Exact SSA (e.g., Gillespie Direct) 
		& Event-by-event simulation 
		& Statistically exact trajectories 
		& High computational cost; scales with reaction firings \cite{gillespie1977,anderson2007} \\
		\addlinespace
		
		Tau-leaping \& variants 
		& Time discretisation; leap over multiple events 
		& Order-of-magnitude speedup for non-stiff systems 
		& Approximation bias; step-size control \cite{gillespie2001,cao2006,anderson2008} \\
		\addlinespace
		
		R-leaping \& batching 
		& Event discretisation; batch multiple firings 
		& Efficient for stiff regimes with bounded populations 
		& Approximation bias; complex sampling logic \cite{rleap1,rleap2} \\
		\addlinespace
		
		Hybrid / multiscale methods 
		& Partition species: discrete (SSA) vs.\ continuous (ODE/CLE) 
		& Speedup when copy numbers vary widely 
		& Heuristic partitioning criteria; potential boundary artefacts \cite{hybrid1,hybrid2} \\
		\addlinespace
		
		\textbf{Reduced precision (this work)} 
			& Reduce memory traffic via 16-bit formats; apply SR 
		& Orthogonal to algorithmic methods; targets memory bottleneck; preserves SSA semantics 
		& Requires careful precision mapping; some operations need \texttt{fp32} protection \\
		\bottomrule
	\end{tabular}
	\caption{Exact SSA baseline and acceleration strategies for stochastic simulation: algorithmic modifications versus reduced precision.}
	\label{tab:taxonomy}
\end{table}

\section{The Stochastic Simulation Algorithm}
\label{sec:ssa}

Consider a well-stirred chemical system of $N$ molecular species $\{X_1, \ldots, X_N\}$ that interact through $M$ reaction channels $\{R_1, \ldots, R_M\}$. The state of the system at time $t$ is the population vector $\mathbf{x}(t) = (x_1, \ldots, x_N)^T$, where $x_i$ is the copy number of species $X_i$. Each reaction channel $R_j$ is characterised by a propensity function $a_j(\mathbf{x})$ and a stoichiometric change vector $\boldsymbol{\nu}_j$: in a small time interval $[t, t+dt)$, the probability that reaction $R_j$ fires is $a_j(\mathbf{x})\,dt$.

The Stochastic Simulation Algorithm (SSA), introduced by Gillespie \cite{gillespie1977}, generates exact realisations of the underlying chemical master equation. In the Direct Method formulation, each step proceeds as follows:
\begin{enumerate}
    \item Compute all propensities $a_j(\mathbf{x})$ for $j = 1, \ldots, M$ and their sum $a_0 = \sum_{j=1}^{M} a_j(\mathbf{x})$,
    \item Draw the waiting time $\tau = -\ln(u_1)/a_0$, where $u_1 \sim \mathrm{Uniform}(0,1)$,
    \item Select the next reaction $R_\mu$ by finding the smallest $\mu$ such that $\sum_{j=1}^{\mu} a_j > u_2 \cdot a_0$, where $u_2 \sim \mathrm{Uniform}(0,1)$,
    \item Update the state $\mathbf{x} \leftarrow \mathbf{x} + \boldsymbol{\nu}_\mu$ and advance time $t \leftarrow t + \tau$.
\end{enumerate}

This procedure generates statistically exact realisations of the continuous-time Markov chain underlying the chemical system, but is computationally expensive: each reaction firing requires a full pass over propensities, two random number draws, and a stoichiometric update of the state vector. For systems with many reactions or fast timescales, the cost of generating long trajectories or large ensembles can be prohibitive. Variants such as the Next Reaction Method \cite{anderson2007}, tau-leaping \cite{gillespie2001,cao2006}, and postleap corrections \cite{anderson2008} offer various trade-offs between exactness and computational efficiency.

From a numerical perspective, three operations in the Direct Method are particularly sensitive to floating-point precision:
\begin{enumerate}
    \item \textbf{Propensity summation.} Computing $a_0 = \sum a_j$ involves additions that may span several orders of magnitude. When fast and slow reactions coexist, catastrophic cancellation or swamping can occur if the accumulator lacks sufficient precision.
    \item \textbf{Logarithmic waiting time.} The transformation $\tau = -\ln(u_1)/a_0$ amplifies errors near $u_1 \approx 0$ or $u_1 \approx 1$, and the division by $a_0$ propagates any error in the propensity sum.
    \item \textbf{Time accumulation.} The running sum $t \leftarrow t + \tau$ is a classical long accumulation: as $t$ grows large relative to $\tau$, low-order bits of $\tau$ may be lost entirely, causing the simulation to stagnate or lose temporal resolution.
\end{enumerate}

\section{Reduced precision and stochastic rounding}
\label{sec:rp_sr}

Digital computers represent real numbers using a finite number of binary digits, so only a discrete subset of the real line is exactly representable. The IEEE~754 standard encodes a floating-point number as $(-1)^s \times 2^{e} \times (1 + f)$, where $s$ is a sign bit, $e$ is an integer exponent that sets the order of magnitude, and $f$ is a fractional mantissa that determines the precision within that range. The number of bits allocated to $e$ and $f$ governs the trade-off between dynamic range and resolution: more exponent bits extend the range of representable magnitudes, while more mantissa bits reduce the spacing between adjacent representable values. Any arithmetic result that falls between two representable numbers must be rounded, and the accumulation of such rounding errors is the central concern when reducing precision. Most scientific computing relies on 64-bit (\texttt{fp64}) or 32-bit (\texttt{fp32}) formats, which provide ample precision for typical applications. However, the proliferation of machine-learning hardware with native support for 16-bit arithmetic creates an opportunity to halve the data footprint of memory-bound algorithms such as SSA, provided the coarser resolution can be managed.

\subsection{Half-precision floating-point formats}

The most widely used reduced-precision format is \texttt{binary16} (commonly called \texttt{fp16}), a 16-bit format with a 1-bit sign, 5-bit exponent, and 10-bit mantissa \cite{ieee754_2019}. This yields a machine epsilon of $\epsilon \approx 9.77 \times 10^{-4}$ and a maximum representable value of 65,504. An alternative 16-bit format, \texttt{bfloat16} (\texttt{bf16}), was developed for deep learning workloads \cite{kalamkar2019}. It uses an 8-bit exponent and 7-bit mantissa, matching the dynamic range of \texttt{fp32} ($\sim 3.4 \times 10^{38}$) at the expense of coarser resolution ($\epsilon \approx 7.81 \times 10^{-3}$).

\begin{table}[ht]
    \centering
    \begin{tabular}{@{}lcccc@{}}
        \toprule
        Format & Exponent bits & Mantissa bits & $\epsilon_{\mathrm{mach}}$ & Max value \\
        \midrule
        \texttt{fp64} & 11 & 52 & $2.2 \times 10^{-16}$ & $1.8 \times 10^{308}$ \\
        \texttt{fp32} & 8 & 23 & $1.2 \times 10^{-7}$ & $3.4 \times 10^{38}$ \\
        \texttt{fp16} & 5 & 10 & $9.8 \times 10^{-4}$ & $6.6 \times 10^{4}$ \\
        \texttt{bf16} & 8 & 7 & $7.8 \times 10^{-3}$ & $3.4 \times 10^{38}$ \\
        \bottomrule
    \end{tabular}
    \caption{Comparison of floating-point formats used in this study. Mantissa bits indicate stored bits, excluding the implicit leading 1; machine epsilon is $\epsilon = 2^{1-p}$, where $p$ is the total precision (stored bits $+\, 1$).}
    \label{tab:formats}
\end{table}

Table~\ref{tab:formats} summarises the key properties of the formats used in this study. The choice between \texttt{fp16} and \texttt{bf16} presents a trade-off: \texttt{fp16} offers finer resolution (10-bit mantissa) but a narrow dynamic range that risks overflow for propensities exceeding $\sim$65,000; \texttt{bf16} tolerates the full \texttt{fp32} dynamic range but rounds more aggressively (7-bit mantissa). Both formats halve memory footprint relative to \texttt{fp32} and are natively supported on modern GPUs and TPUs \cite{jouppi2017,kalamkar2019}.

\subsection{Mixed-precision partitioning}

The numerical sensitivities identified in \Cref{sec:ssa} suggest a natural partitioning of SSA operations by precision requirement. Propensity calculations involve multiplications and additions of rate constants and population counts, operations that are typically well-conditioned when populations are moderate. By contrast, the propensity sum $a_0$, the waiting-time logarithm, and time accumulation demand higher precision to avoid systematic error.

This suggests a mixed-precision strategy: compute propensities in a reduced-precision type \texttt{Tprop} $\in \{\texttt{fp16}, \texttt{bf16}\}$ but promote to a higher-precision accumulator type \texttt{Tacc} $=$ \texttt{fp32} before summing propensities and computing derived quantities. Time is maintained in \texttt{fp64} to prevent stagnation during long simulations. In Julia's type system, this partitioning is implemented by parameterising the SSA kernel on two type parameters \texttt{Tprop} and \texttt{Tacc}, allowing precision choices to be varied without altering control flow. State variables (population counts) are maintained as native integers throughout the simulation; only propensity evaluation operates in \texttt{Tprop}. Stoichiometric updates ($\pm 1$, $\pm 2$, etc.) are therefore exact integer arithmetic, independent of the precision used for propensities.

We also evaluate a ``uniform'' mode in which all arithmetic (including accumulation and time) uses the reduced-precision type. This is intentionally aggressive: it probes the limits of low-precision SSA and reveals which operations truly require protection.

\subsection{Stochastic rounding}

Standard IEEE rounding (round-to-nearest, RTN) is deterministic: a value $x$ falling between two representable numbers is always mapped to the nearer one. While this minimises worst-case error for individual operations, it introduces systematic bias when small increments are repeatedly added to a large accumulator. If the increment is smaller than half the spacing between adjacent representables at the accumulator's magnitude, RTN rounds it to zero every time, and the accumulator stagnates.

SR \cite{parker1997} replaces this deterministic map with a probabilistic one. For a real value $x$ lying between adjacent representable numbers $x^{-}$ and $x^{+}$, SR returns
\begin{equation}
    \mathrm{SR}(x) = \begin{cases}
        x^{-} & \text{with probability } \frac{x^{+} - x}{x^{+} - x^{-}}, \\[4pt]
        x^{+} & \text{with probability } \frac{x - x^{-}}{x^{+} - x^{-}}.
    \end{cases}
\end{equation}
The key property is unbiasedness: $\mathbb{E}[\mathrm{SR}(x)] = x$. Over many additions, SR errors behave as a zero-mean random walk rather than accumulating systematically, preventing the stagnation that plagues RTN in long reductions \cite{connolly2021,croci2022}.

For SSA, SR is particularly relevant in two contexts. First, time accumulation $t \leftarrow t + \tau$ in uniform reduced-precision mode: when $t \gg \tau$, RTN may round $\tau$ away entirely, causing the simulation clock to freeze. SR ensures that small time increments contribute on average, maintaining temporal progress. Second, propensity sums involving reactions with vastly different rates: SR prevents slow-reaction propensities from being systematically rounded to zero when added to fast-reaction propensities.

In this study, we use the \texttt{StochasticRounding.jl} Julia package, which provides drop-in \texttt{Float16sr} and \texttt{BFloat16sr} types that apply SR at every arithmetic operation. This ``every-op'' SR strategy is the most conservative; selective SR (applied only at accumulation points) may suffice in practice but is not explored here. Previous work has demonstrated the effectiveness of SR for maintaining long-term statistical accuracy in climate models \cite{kimpson2024} and numerical PDE solvers \cite{croci2022}.

\section{Numerical Experiments}
\label{sec:experiments}

We evaluate mixed-precision SSA on five canonical biochemical kinetics models, chosen to stress different numerical regimes. For each model, we compare six precision configurations:
\begin{itemize}
    \item \textbf{FP64}: 64-bit floating point (reference baseline)
    \item \textbf{FP32}: 32-bit floating point
    \item \textbf{FP16 RTN}: IEEE half precision with round-to-nearest
    \item \textbf{FP16+SR}: IEEE half precision with SR
    \item \textbf{BF16 RTN}: Brain float with round-to-nearest
    \item \textbf{BF16+SR}: Brain float with SR
\end{itemize}
In ``mixed precision'' mode, propensity calculations use reduced precision while time and state accumulators remain in FP32. In ``uniform'' mode, all operations (including accumulation and time) use the reduced-precision format. We run ensembles of 50,000 independent trajectories for each model and precision configuration, except for the Schl\"{o}gl model (10,000 trajectories) and the repressilator (1,000 trajectories), where longer simulation times and higher per-trajectory costs make larger ensembles impractical. Rate constants throughout are reported in arbitrary units of inverse time; the steady-state distributions analysed here are invariant under a global rescaling of time.

We compare each reduced-precision ensemble against the FP64 reference using two complementary metrics. The two-sample Kolmogorov--Smirnov (KS) test \cite{kolmogorov1933,smirnov1948} assesses whether two samples are drawn from the same distribution. Here, each sample is the ensemble of per-trajectory state values at the simulation end time, and we compare the FP64 ensemble against each reduced-precision ensemble: a $p$-value near 1 indicates no detectable difference, while $p < 0.05$ rejects the null hypothesis of distributional equivalence. The Wasserstein-1 distance (earth mover's distance) \cite{vaserstein1969,villani2009} quantifies the average absolute displacement needed to transform one distribution into the other; it inherits the units of the distributions being compared (here, molecules) and is zero if and only if the distributions are identical. We additionally report sample means and variances of the steady-state distributions.

\subsection{Birth--Death Process}

\begin{figure}[ht]
    \centering
    \begin{subfigure}[t]{0.48\textwidth}
        \centering
        \includegraphics[width=\textwidth]{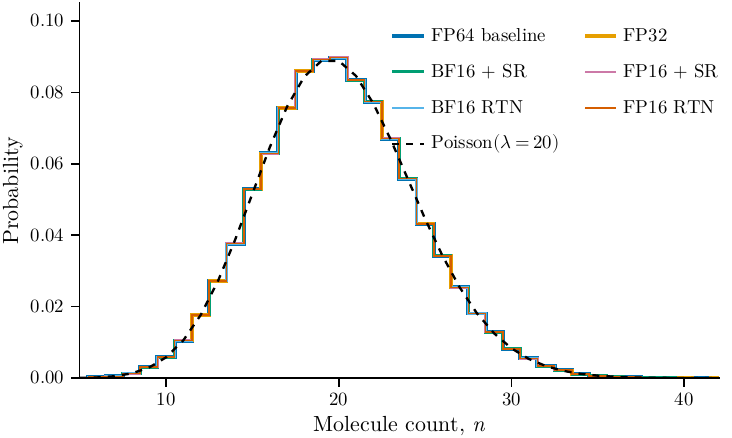}
        \caption{Mixed precision}
    \end{subfigure}
    \hfill
    \begin{subfigure}[t]{0.48\textwidth}
        \centering
        \includegraphics[width=\textwidth]{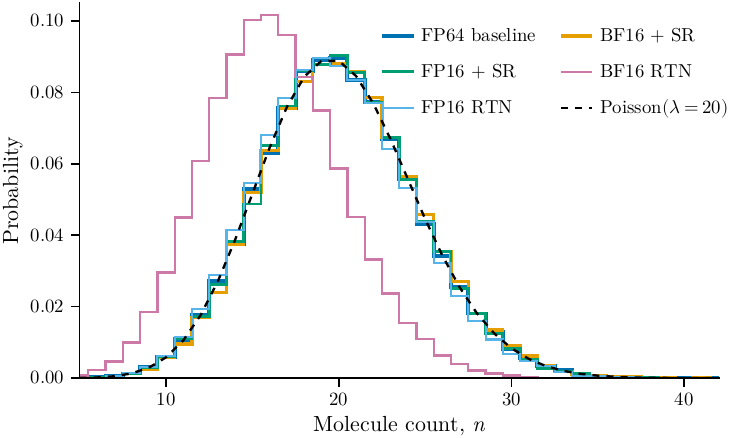}
        \caption{Uniform precision}
    \end{subfigure}
    \caption{Steady-state distribution of the birth--death process ($N = 50{,}000$ realisations each), compared to the analytical Poisson distribution (black dashed line). (a) All six mixed-precision configurations produce indistinguishable results. (b) Uniform-mode comparison: SR modes closely match the FP64 reference, while BF16 RTN (pink) shifts left to mean $\approx 16$, where time stagnation terminates the simulation prematurely.}
    \label{fig:birth-death}
\end{figure}

The birth--death process \cite{gillespie1977} is a minimal linear kinetics model with reactions
\begin{align*}
    \varnothing &\xrightarrow{k_b} X, \quad X \xrightarrow{k_d} \varnothing \, ,
\end{align*}
where we take $k_b = 20$ and $k_d = 1$, yielding a Poisson steady-state distribution with mean $\lambda = k_b/k_d = 20$. We use it as a baseline for well-conditioned numerical problems. As shown in Table~\ref{tab:birth-death} and Figure~\ref{fig:birth-death}, all six mixed-precision modes achieved perfect agreement with the analytical solution. Linear kinetics with moderate population counts ($\sim$20) were robust to precision reduction in mixed mode. Reduced precision did not artificially degrade results when the numerical problem was well-conditioned and sensitive operations were protected by FP32 accumulators.

\begin{table}[ht]
    \centering
    \begin{tabular}{@{}lcccc@{}}
        \toprule
        Precision Mode & KS $p$-value & Wasserstein & Mean & Variance \\
        \midrule
        \multicolumn{5}{l}{\textit{Mixed precision}} \\
        \midrule
        FP64 (baseline) & 1.0 & 0 & 20.00 & 20.04 \\
        FP32 & 1.0 & 0 & 20.00 & 20.04 \\
        FP16 RTN & 1.0 & 0 & 20.00 & 20.04 \\
        FP16 + SR & 1.0 & 0 & 20.00 & 20.04 \\
        BF16 RTN & 1.0 & 0 & 20.00 & 20.04 \\
        BF16 + SR & 1.0 & 0 & 20.00 & 20.04 \\
        \midrule
        \multicolumn{5}{l}{\textit{Uniform precision}} \\
        \midrule
        FP16 RTN & $2.3 \times 10^{-8}$ & 0.22 & 19.78 & 19.56 \\
        FP16 + SR & 0.681 & 0.03 & 20.01 & 19.77 \\
        \textbf{BF16 RTN} & $\mathbf{\approx 0}$ & \textbf{3.75} & \textbf{16.25} & \textbf{15.84} \\
        BF16 + SR & 0.001 & 0.12 & 20.12 & 19.98 \\
        \bottomrule
    \end{tabular}
    \caption{Results for the birth--death model: mixed-precision modes achieve perfect agreement with the analytical Poisson distribution. In uniform mode, BF16 RTN fails due to time stagnation. Bold entries indicate qualitative failure (distributional corruption or overflow).}
    \label{tab:birth-death}
\end{table}

In uniform precision mode, where all operations including accumulation and time used reduced precision, failures emerged (Table~\ref{tab:birth-death}). Uniform BF16 RTN failed even on this simplest of models: the mean dropped from $20.0$ to $16.25$ (Wasserstein $= 3.75$, KS $p \approx 0$). The failure mechanism is time stagnation. Each step samples a valid time increment $\tau$, but BF16's 7-bit mantissa causes the addition $t \leftarrow t + \tau$ to round back to $t$ when the accumulated simulation time greatly exceeds $\tau$. Reactions continue to fire and the state evolves, but the simulated clock barely advances; the simulation then terminates before reaching equilibration, yielding a state biased toward initial conditions. Uniform FP16 RTN showed a statistically significant but modest shift (mean $= 19.78$, KS $p = 2.3 \times 10^{-8}$). Turning to SR, uniform FP16+SR maintained excellent agreement with the baseline (Wasserstein $= 0.03$, KS $p = 0.68$), while uniform BF16+SR showed a marginal but detectable shift (mean $20.12$, KS $p = 0.001$).

\subsection{Schl\"{o}gl Model}

The Schl\"{o}gl model \cite{schlogl1972} exhibits bistability through autocatalytic reactions:
\begin{align*}
    A + 2X &\underset{k_2}{\overset{k_1}{\rightleftharpoons}} 3X, \quad B \underset{k_4}{\overset{k_3}{\rightleftharpoons}} X.
\end{align*}
Its bistability and rare-event dynamics provide a stringent test for precision effects. We use the classic bistable parameterisation \cite{hybrid2} with rate constants that absorb combinatorial prefactors: $k_1 = 1.5\times10^{-7}$, $k_2 = 1.667\times10^{-5}$, $k_3 = 10^{-3}$, $k_4 = 3.5$, and reservoir copy numbers $A = 10^5$, $B = 2\times10^5$, yielding a bimodal stationary distribution with modes near $n_X \approx 90$ and $n_X \approx 560$, where $n_X$ denotes the copy number of $X$.

\begin{figure}[ht]
    \centering
    \begin{subfigure}[t]{0.48\textwidth}
        \centering
        \includegraphics[width=\textwidth]{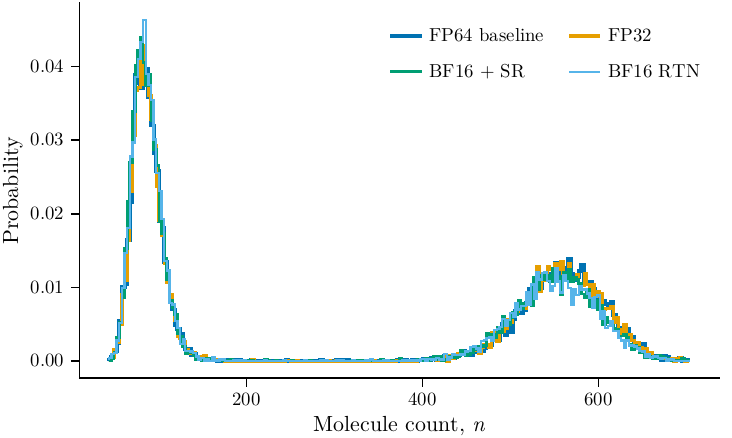}
        \caption{Original --- mixed precision}
    \end{subfigure}
    \hfill
    \begin{subfigure}[t]{0.48\textwidth}
        \centering
        \includegraphics[width=\textwidth]{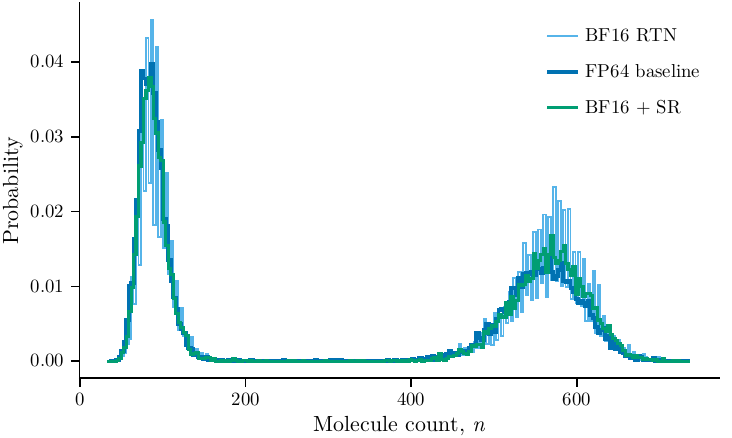}
        \caption{Original --- uniform precision}
    \end{subfigure}

    \medskip

    \begin{subfigure}[t]{0.48\textwidth}
        \centering
        \includegraphics[width=\textwidth]{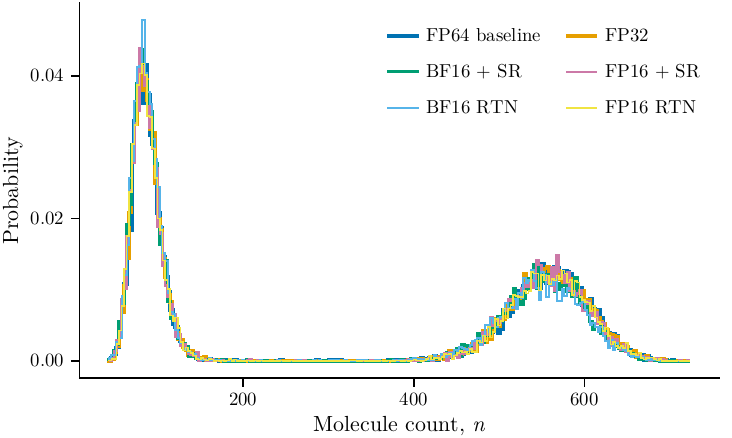}
        \caption{Rescaled --- mixed precision}
    \end{subfigure}
    \hfill
    \begin{subfigure}[t]{0.48\textwidth}
        \centering
        \includegraphics[width=\textwidth]{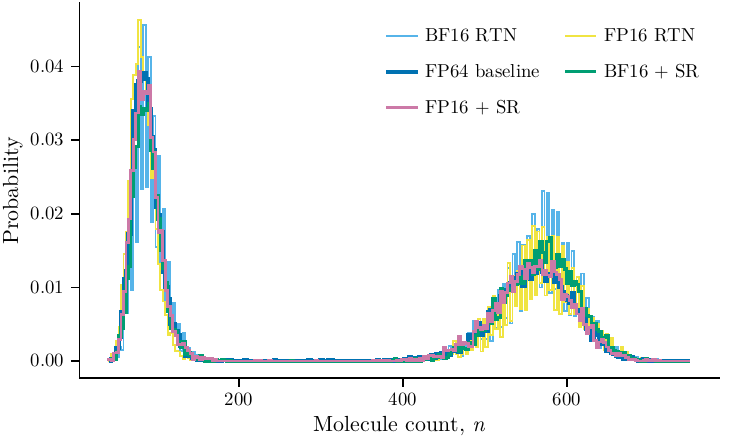}
        \caption{Rescaled --- uniform precision}
    \end{subfigure}
    \caption{Bimodal stationary distribution of the Schl\"{o}gl system ($N = 10{,}000$ realisations each). Top row: original parameterisation. Bottom row: rescaled parameterisation.
    (a)~FP32 is indistinguishable from FP64; BF16 modes show minor shifts in peak occupancy. FP16 modes are excluded (cubic propensity overflow).
    (b)~BF16+SR slightly favours the high peak; BF16 RTN shifts further toward the high peak. FP16 uniform modes overflow identically to their mixed counterparts.
    (c)~All six modes reproduce the bimodal structure; FP16+SR is nearly indistinguishable from FP64, confirming the original failure was due to parameter range.
    (d)~FP16+SR ($W_1 = 2.76$, KS $p = 0.81$) closely matches FP64, while BF16 uniform modes shift toward the high peak.}
    \label{fig:schlogl}
\end{figure}

In mixed-precision mode (Table~\ref{tab:schlogl}, Figure~\ref{fig:schlogl}(a)), FP32 closely reproduced the FP64 bimodal distribution ($W_1 = 1.41$), and BF16 modes showed moderate deviations ($W_1 \approx 20$--$24$), manifesting as shifts in relative peak occupancy rather than corruption of the bimodal structure itself. FP16 modes (both SR and RTN), however, failed due to dynamic range: the cubic propensity $k_2 \,  n(n{-}1)(n{-}2)$ reaches $\sim\!10^7$ at the high peak ($n \approx 560$), far exceeding FP16's maximum representable value of $65{,}504$. No rounding mode can compensate for this hard overflow. BF16, by contrast, shares FP32's exponent range ($\sim\!3.4\times10^{38}$), so propensity values posed no range problem. These results demonstrated that mixed-precision SSA with BF16+SR successfully captured bistable dynamics.

Uniform precision simulations amplified the effect of mantissa resolution (Figure~\ref{fig:schlogl}(b)). BF16's 7-bit mantissa introduces $\sim$0.3\% relative error per operation; under RTN, these errors are systematically directional. A plausible mechanism is that in the four-multiplication propensity chain $k_2 \, n \, (n{-}1) \, (n{-}2)$, compounding bias skews reaction selection toward population growth, which increases the propensity magnitude, and amplifies the bias further. Uniform BF16 RTN ($W_1 = 35.9$, mean $= 353.1$) shifted the distribution substantially toward the high peak. SR eliminated this feedback loop: each rounding error is zero-mean, so biases cancel statistically rather than accumulating. This made uniform BF16+SR the only viable uniform-mode configuration for the Schl\"{o}gl system, though even it showed moderate deviation ($W_1 = 22.9$, mean $= 340.1$). FP16 uniform modes overflowed identically to their mixed counterparts.

\begin{table}[ht]
    \centering
    \begin{tabular}{@{}lcccc@{}}
        \toprule
        Precision Mode & KS $p$-value & Wasserstein & Mean & Variance \\
        \midrule
        \multicolumn{5}{l}{\textit{Mixed precision}} \\
        \midrule
        FP64 (baseline)   & 1.0    & 0      & 317.2 & 56674 \\
        FP32              & 0.834  & 1.41   & 315.9 & 56459 \\
        \textbf{FP16 RTN} & \textbf{overflow} & \textbf{233.7} & \textbf{250.0} & \textbf{0} \\
        \textbf{FP16 + SR} & \textbf{overflow} & \textbf{233.7} & \textbf{250.0} & \textbf{0} \\
        BF16 RTN          & $<$0.001 & 24.1 & 293.1 & 54107 \\
        BF16 + SR         & $<$0.001 & 19.7 & 297.5 & 54816 \\
        \midrule
        \multicolumn{5}{l}{\textit{Uniform precision}} \\
        \midrule
        \textbf{FP16} & \textbf{overflow} & \textbf{233.7} & \textbf{250.0} & \textbf{0} \\
        \textbf{BF16 RTN} & $\mathbf{\approx 0}$ & \textbf{35.9} & \textbf{353.1} & \textbf{57385} \\
        BF16 + SR  & $<$0.001 & 22.9 & 340.1 & 57956 \\
        \bottomrule
    \end{tabular}
    \caption{Results for the Schl\"{o}gl model: precision comparison ($N = 10{,}000$ realisations, $t_{\mathrm{end}} = 10$). FP32 closely matches FP64; BF16 modes show moderate deviation. FP16 modes overflow (mean clamped at 250). All uniform FP16 modes (SR and RTN) overflow identically. Uniform BF16 RTN diverges. Bold entries indicate qualitative failure (distributional corruption or overflow).}
    \label{tab:schlogl}
\end{table}

The FP16 exclusion above is a parameter representation limitation, not a fundamental arithmetic precision problem. The SSA dynamics depend only on the products $k_1 A$ and $k_3 B$, not on the individual values. By reparameterising (setting $k_1 = 1.5\times10^{-4}$, $A = 100$, $k_3 = 1.0$, $B = 200$ while preserving $k_1 A = 0.015$ and $k_3 B = 200$), all parameter values and intermediate propensities remain within FP16's representable range (verified: $a_1 = 4696$, $a_2 = 2914$, $a_3 = 200$, $a_4 = 1960$ at $n = 560$, all $\ll 65{,}504$). This nondimensionalisation preserves identical peak positions and transition dynamics.

\begin{table}[ht]
    \centering
    \begin{tabular}{@{}lcccc@{}}
        \toprule
        Precision Mode & KS $p$-value & Wasserstein & Mean & Variance \\
        \midrule
        \multicolumn{5}{l}{\textit{Mixed precision}} \\
        \midrule
        FP64 (baseline)   & 1.0    & 0      & 317.2 & 56674 \\
        FP32              & 0.834  & 1.41   & 315.9 & 56459 \\
        FP16 RTN          & 0.041  & 7.52   & 309.8 & 56253 \\
        FP16 + SR         & 0.040  & 3.92   & 313.4 & 56005 \\
        BF16 RTN          & $<$0.001 & 24.1 & 293.1 & 54107 \\
        BF16 + SR         & $<$0.001 & 15.8 & 301.5 & 54957 \\
        \midrule
        \multicolumn{5}{l}{\textit{Uniform precision}} \\
        \midrule
        \textbf{FP16 RTN} & $\mathbf{\approx 0}$ & \textbf{8.95} & \textbf{322.3} & \textbf{59744} \\
        FP16 + SR  & 0.813  & 2.76   & 319.8 & 56557 \\
        \textbf{BF16 RTN} & $\mathbf{\approx 0}$ & \textbf{35.9} & \textbf{353.1} & \textbf{57385} \\
        BF16 + SR  & $<$0.001 & 31.1 & 348.4 & 58024 \\
        \bottomrule
    \end{tabular}
    \caption{Results for the rescaled Schl\"{o}gl model: mixed and uniform precision modes ($N = 10{,}000$ realisations, $t_{\mathrm{end}} = 10$). FP16 modes are now viable after reparameterisation. Uniform FP16+SR closely matches FP64; uniform BF16 modes show large deviations. Bold entries indicate qualitative failure (distributional corruption or overflow).}
    \label{tab:schlogl-rescaled}
\end{table}

Table~\ref{tab:schlogl-rescaled} and Figure~\ref{fig:schlogl}(c) confirmed that FP16 became fully viable after rescaling. In mixed-precision mode, FP16+SR achieved a Wasserstein distance of $3.92$, dramatically better than BF16+SR ($15.8$). FP16 RTN also performed well (Wasserstein $= 7.52$), outperforming both BF16 modes. In uniform mode (Figure~\ref{fig:schlogl}(d)), rescaling unlocked uniform FP16+SR, which achieved $W_1 = 2.76$ with KS $p = 0.81$, passing both validation criteria comfortably. This surpassed the mixed FP16+SR result ($W_1 = 3.92$). The original FP16 failure was entirely due to parameter range, not arithmetic precision. Uniform BF16 modes remained problematic ($W_1 = 31.1$ for SR, $35.9$ for RTN), reflecting BF16's coarser 7-bit mantissa.

More broadly, this experiment illustrates a practical point: observing failure when switching to reduced precision does not mean the underlying system cannot be faithfully represented at that precision. The original Schl\"{o}gl parameterisation overflows in FP16 not because half-precision arithmetic is too coarse for bistable dynamics, but because specific parameter values ($A = 10^5$) exceed the format's representable range. Standard nondimensionalisation, a routine step in computational physics, eliminates this obstruction entirely. Kl\"{o}wer et al.~\cite{Klower2022} make an analogous observation in the context of fluid dynamics: rescaling state variables and physical constants to avoid overflow and underflow is often sufficient to unlock the performance benefits of 16-bit arithmetic, and failure in a na\"{i}ve precision reduction is not necessarily evidence of fundamental incompatibility.

\subsection{Telegraph Model}

The Telegraph model \cite{peccoud1995} describes bursty gene expression with stochastic promoter switching:
\begin{align*}
    G_{\text{off}} &\underset{k_{\mathrm{off}}}{\overset{k_{\mathrm{on}}}{\rightleftharpoons}} G_{\text{on}}, \quad G_{\text{on}} \xrightarrow{\alpha} G_{\text{on}} + M, \quad M \xrightarrow{\beta} \varnothing.
\end{align*}
Here $\alpha$ is the transcription rate and $\beta$ the mRNA degradation rate, and we write $n_M$ for the mRNA copy number. The steady-state distribution of $n_M$ is governed by the dimensionless ratios $a = k_{\mathrm{on}}/\beta$, $b = k_{\mathrm{off}}/\beta$, and $\lambda = \alpha/\beta$. Depending on these parameters, the model exhibits qualitatively different behaviour, ranging from unimodal Poisson-like distributions when switching is fast ($a, b \gg 1$) to bimodal distributions reflecting the two promoter states when switching is slow ($a, b \ll 1$). We examine two slow-switching parameterisations that probe different aspects of precision sensitivity.

We first consider $k_{\mathrm{on}} = k_{\mathrm{off}} = 0.05$, $\alpha = 20$, $\beta = 1$, giving dimensionless parameters $a = b = 0.05$ and $\lambda = 20$. The gene spends equal time in the ON and OFF states, and the slow switching produces a bimodal mRNA distribution with mean $\approx 10$. Because mRNA decay ($\beta = 1$) is much faster than promoter switching ($k_{\mathrm{on}} = 0.05$), the molecule count drops to zero well before the gene reactivates, producing a sharp spike at $n_M = 0$ containing $\sim$42\% of the probability mass. The remaining mass forms a broad peak near $\lambda = 20$, corresponding to the ON-state burst.

\begin{figure}[ht]
    \centering
    \begin{subfigure}[t]{0.48\textwidth}
        \centering
        \includegraphics[width=\textwidth]{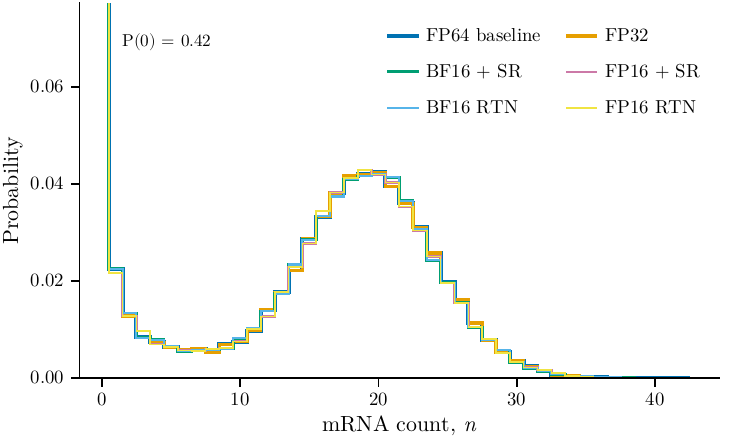}
        \caption{Bimodal --- mixed precision}
    \end{subfigure}
    \hfill
    \begin{subfigure}[t]{0.48\textwidth}
        \centering
        \includegraphics[width=\textwidth]{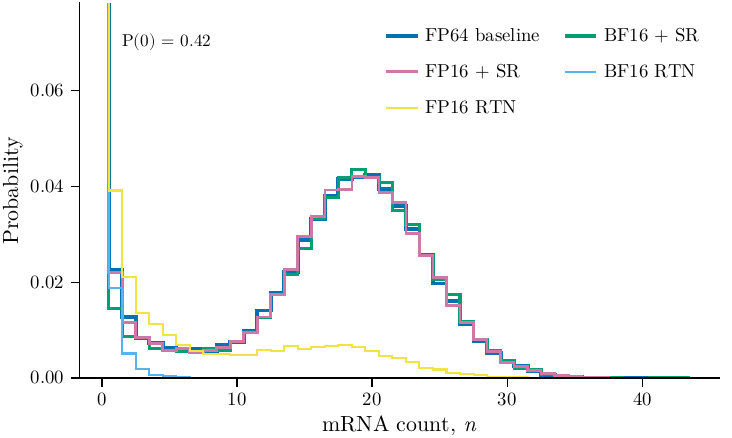}
        \caption{Bimodal --- uniform precision}
    \end{subfigure}

    \medskip

    \begin{subfigure}[t]{0.48\textwidth}
        \centering
        \includegraphics[width=\textwidth]{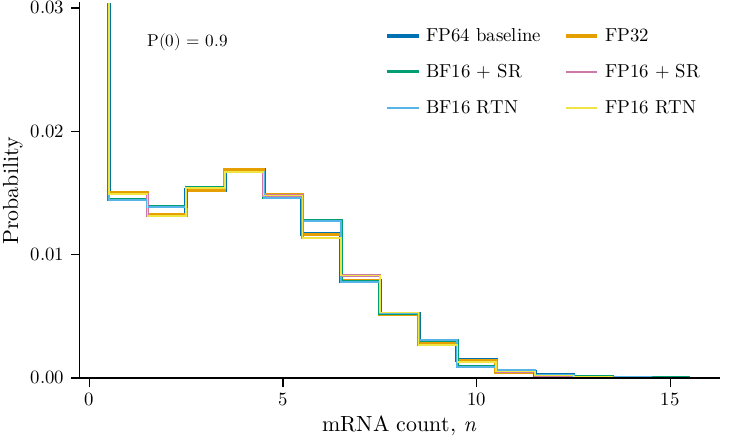}
        \caption{Underflow stress --- mixed precision}
    \end{subfigure}
    \hfill
    \begin{subfigure}[t]{0.48\textwidth}
        \centering
        \includegraphics[width=\textwidth]{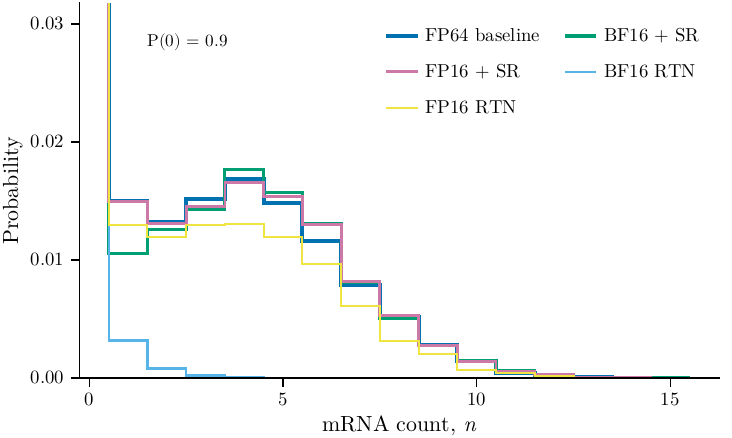}
        \caption{Underflow stress --- uniform precision}
    \end{subfigure}
    \caption{Telegraph mRNA marginal distributions. Top row: bimodal parameterisation ($y$-axis truncated; $P(n{=}0) = 0.42$). Bottom row: underflow stress parameterisation ($k_{\mathrm{on}} = 0.01$; $y$-axis truncated; $P(n{=}0) = 0.90$).
    (a)~Mixed-precision modes faithfully reproduce the two-peaked structure.
    (b)~SR modes match the FP64 reference, while BF16 RTN collapses entirely to zero and FP16 RTN is severely degraded.
    (c)~Mixed-precision modes all reproduce the geometric-like tail.
    (d)~SR modes match the FP64 reference, while BF16 RTN collapses to $n_M = 0$; the gene remains perpetually off.}
    \label{fig:telegraph-marginals}
\end{figure}

As shown in Table~\ref{tab:telegraph-bimodal} and Figure~\ref{fig:telegraph-marginals}(a), mixed-precision mode reproduced the bimodal structure without difficulty. All six configurations achieved excellent agreement with the FP64 baseline, with Wasserstein distances below 0.07 and KS $p$-values exceeding 0.86. In uniform precision mode (Table~\ref{tab:telegraph-bimodal} and Figure~\ref{fig:telegraph-marginals}(b)), the same SR-vs-RTN dichotomy emerged. Uniform FP16+SR maintained excellent agreement (Wasserstein $= 0.041$, KS $p = 0.83$), while uniform BF16+SR showed a marginal KS $p$-value of $4.3 \times 10^{-4}$ but preserved the bimodal shape with Wasserstein $= 0.107$. Without SR, both formats failed: uniform BF16 RTN collapsed to mean $= 0.04$ (Wasserstein $= 10.0$), and uniform FP16 RTN dropped to mean $= 1.84$ (Wasserstein $= 8.2$). The failure mechanism is the same as in the original parameterisation: $k_{\mathrm{on}} = 0.05$ is swamped by the total propensity ($a_0 \approx 40$ when the gene is ON with $\sim$20 mRNA molecules) under RTN, preventing the OFF$\to$ON transition from being selected.

\begin{table}[ht]
    \centering
    \begin{tabular}{@{}lcccc@{}}
        \toprule
        Precision Mode & KS $p$-value & Wasserstein & Mean & Variance \\
        \midrule
        \multicolumn{5}{l}{\textit{Mixed precision}} \\
        \midrule
        FP64 (baseline) & 1.0 & 0 & 10.05 & 100.9 \\
        FP32 & 1.0 & 0 & 10.05 & 100.9 \\
        FP16 RTN & 0.863 & 0.069 & 9.98 & 100.8 \\
        FP16 + SR & 0.863 & 0.069 & 9.98 & 100.8 \\
        BF16 RTN & 0.966 & 0.052 & 10.00 & 100.7 \\
        BF16 + SR & 0.966 & 0.052 & 10.00 & 100.7 \\
        \midrule
        \multicolumn{5}{l}{\textit{Uniform precision}} \\
        \midrule
        \textbf{FP16 RTN} & $\mathbf{\approx 0}$ & \textbf{8.20} & \textbf{1.84} & \textbf{25.2} \\
        FP16 + SR & 0.828 & 0.041 & 10.03 & 101.5 \\
        \textbf{BF16 RTN} & $\mathbf{\approx 0}$ & \textbf{10.01} & \textbf{0.040} & \textbf{0.080} \\
        BF16 + SR & $4.3 \times 10^{-4}$ & 0.107 & 10.11 & 102.8 \\
        \bottomrule
    \end{tabular}
    \caption{Results for the Telegraph model (bimodal): mRNA steady-state statistics. Mixed-precision modes achieve excellent agreement; uniform SR preserves the bimodal distribution while RTN collapses it. Bold entries indicate qualitative failure (distributional corruption or overflow).}
    \label{tab:telegraph-bimodal}
\end{table}

\begin{figure}[ht]
    \centering
    \begin{subfigure}[t]{0.48\textwidth}
        \centering
        \includegraphics[width=\textwidth]{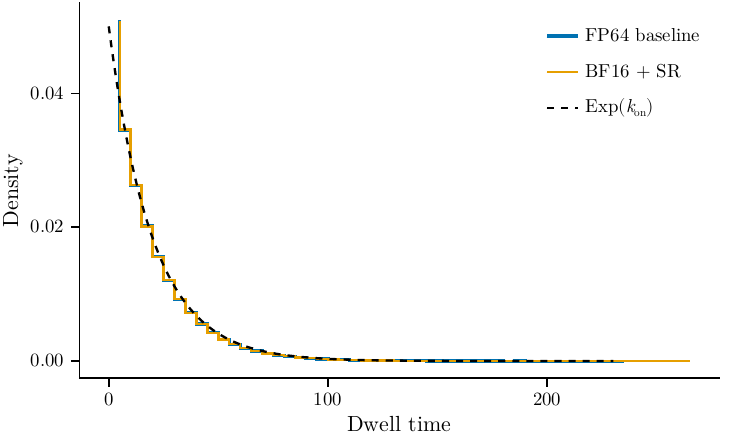}
        \caption{Dwell-time distribution}
    \end{subfigure}
    \hfill
    \begin{subfigure}[t]{0.48\textwidth}
        \centering
        \includegraphics[width=\textwidth]{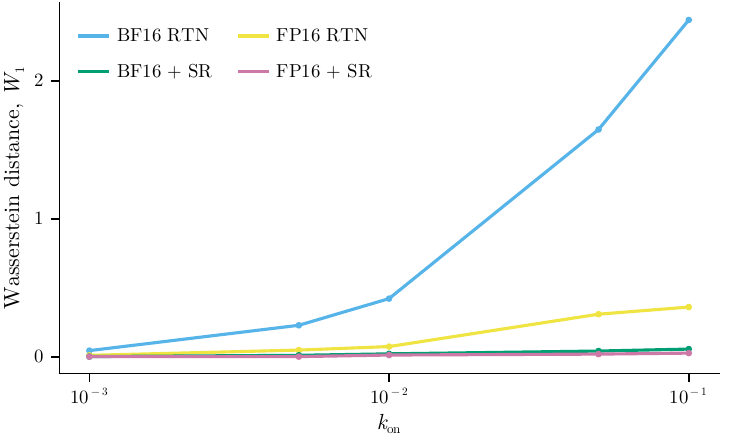}
        \caption{Parameter sweep}
    \end{subfigure}
    \caption{Telegraph model validation.
    (a)~OFF-state dwell-time distribution for the bimodal parameterisation ($k_{\mathrm{on}} = k_{\mathrm{off}} = 0.05$); the ON-state distribution is identical by symmetry. Both FP64 and BF16+SR match the analytical exponential PDF.
    (b)~Uniform-mode parameter sweep over $k_{\mathrm{on}}\in[10^{-3},10^{-1}]$: Wasserstein distance to FP64. SR modes remain near zero across the entire sweep, while RTN modes degrade sharply as $k_{\mathrm{on}}$ decreases.}
    \label{fig:telegraph-validation}
\end{figure}

Independent confirmation of correct switching dynamics is provided by the dwell-time distributions (Figure~\ref{fig:telegraph-validation}(a)). Because $k_{\mathrm{on}} = k_{\mathrm{off}}$, the ON and OFF dwell-time distributions are identical; both FP64 and BF16+SR match the analytical exponential PDF. We next consider a parameterisation designed to maximally stress reduced-precision arithmetic: $k_{\mathrm{on}} = 0.01$, $k_{\mathrm{off}} = 0.1$, $\alpha = 5$, $\beta = 1$, a slow-switching regime with low mRNA copy numbers (mean $\approx 0.44$). The small switching rate $k_{\mathrm{on}} = 0.01$ risks underflow in low-precision formats, promoter transitions are rare relative to mRNA dynamics, and the resulting multi-timescale separation amplifies any systematic rounding bias.

In mixed-precision mode, all six configurations achieved excellent agreement (Table~\ref{tab:telegraph} and Figure~\ref{fig:telegraph-marginals}(c)), with Wasserstein distances below 0.01 and KS $p$-values indistinguishable from unity. Uniform precision mode sharpened the SR-vs-RTN contrast (Table~\ref{tab:telegraph} and Figure~\ref{fig:telegraph-marginals}(d)). With SR enabled, both FP16 and BF16 maintained full statistical validity: KS $p$-values exceeded 0.94 and Wasserstein distances remained near 0.01. Without SR, both formats failed. Uniform FP16 RTN produced a KS $p$-value of $7.8 \times 10^{-9}$ and a Wasserstein distance of 0.098. Uniform BF16 RTN suffered catastrophic collapse: the mean dropped from 0.44 to 0.006 and the KS $p$-value is effectively zero ($4.2 \times 10^{-220}$). BF16's coarse 7-bit mantissa causes $k_{\mathrm{on}} = 0.01$ to be systematically rounded to zero under RTN, so the gene never switches on and mRNA production ceases almost entirely. SR prevented this failure by ensuring that $k_{\mathrm{on}}$ contributes on average, even when it falls below the local representable threshold. The two parameterisations above probe specific $k_{\mathrm{on}}$ values; Figure~\ref{fig:telegraph-validation}(b) generalises across a two-order-of-magnitude $k_{\mathrm{on}}$ sweep ($10^{-3}$ to $10^{-1}$). SR modes remained robust across the entire parameter range, while RTN modes degraded sharply as $k_{\mathrm{on}}$ decreases and the underflow risk intensifies. Across both parameterisations, the Telegraph model confirmed that SR was both necessary and sufficient for maintaining statistical fidelity in uniform reduced-precision SSA on this class of switching model (the repressilator, Section~\ref{sec:repressilator}, will show that sufficiency does not hold for all systems).

\begin{table}[ht]
    \centering
    \begin{tabular}{@{}lcccc@{}}
        \toprule
        Precision Mode & KS $p$-value & Wasserstein & Mean & Variance \\
        \midrule
        \multicolumn{5}{l}{\textit{Mixed precision}} \\
        \midrule
        FP64 (baseline) & 1.0 & 0 & 0.444 & 2.26 \\
        FP32 & 1.0 & 0 & 0.444 & 2.26 \\
        FP16 RTN & 1.0 & 0.0014 & 0.444 & 2.26 \\
        FP16 + SR & 1.0 & 0.0014 & 0.444 & 2.26 \\
        BF16 RTN & 0.9999 & 0.0069 & 0.451 & 2.30 \\
        BF16 + SR & 0.9999 & 0.0069 & 0.451 & 2.30 \\
        \midrule
        \multicolumn{5}{l}{\textit{Uniform precision}} \\
        \midrule
        \textbf{FP16 RTN} & $\mathbf{7.8 \times 10^{-9}}$ & \textbf{0.098} & 0.346 & 1.73 \\
        FP16 + SR & 0.999 & 0.0113 & 0.456 & 2.34 \\
        \textbf{BF16 RTN} & $\mathbf{4.2 \times 10^{-220}}$ & \textbf{0.438} & \textbf{0.006} & \textbf{0.012} \\
        BF16 + SR & 0.943 & 0.0156 & 0.455 & 2.35 \\
        \bottomrule
    \end{tabular}
    \caption{Results for the Telegraph model (underflow stress): mixed precision achieves excellent agreement; uniform SR maintains fidelity while RTN fails catastrophically. Bold entries indicate qualitative failure (distributional corruption or overflow).}
    \label{tab:telegraph}
\end{table}

\subsection{Dimerisation}

The dimerisation model \cite{gillespie2007} describes reversible protein dimerisation,
\begin{align*}
    2A &\underset{k_r}{\overset{k_f}{\rightleftharpoons}} D,
\end{align*}
with forward rate $k_f = 10^{-3}$ and reverse rate $k_r = 0.1$, starting from $n_A(0) = 100$ monomers and $n_D(0) = 0$ dimers. The forward (binding) propensity $a_1 = k_f \, n_A(n_A - 1)/2$ is nonlinear and involves a combinatorial factor counting distinct monomer pairs, while the reverse (unbinding) propensity $a_2 = k_r \, n_D$ is linear. Because each binding event consumes two monomers and produces one dimer, and vice versa, the system obeys the strict conservation law $n_A + 2n_D = 100$ at all times. We simulate both species independently rather than tracking only one and inferring the other; this preserves the dimer model as a faithful test of whether conservation emerges from stoichiometric updates under reduced-precision arithmetic. At steady state the distribution is unimodal, centered near $\langle n_A \rangle \approx 62$ and $\langle n_D \rangle \approx 19$. The combination of a bimolecular propensity and a structural conservation constraint makes this a useful test of reduced-precision arithmetic on coupled species.

\begin{table}[ht]
    \centering
    \begin{tabular}{@{}lcccc@{}}
        \toprule
        Precision Mode & $A$ KS $p$-value & $A$ Wasserstein & $D$ KS $p$-value & $D$ Wasserstein \\
        \midrule
        \multicolumn{5}{l}{\textit{Mixed precision}} \\
        \midrule
        FP64 (baseline) & 1.0 & 0 & 1.0 & 0 \\
        FP32 & 1.0 & 0.001 & 1.0 & 0.0005 \\
        FP16 RTN & 1.0 & 0.016 & 1.0 & 0.008 \\
        FP16 + SR & 0.544 & 0.046 & 0.544 & 0.023 \\
        BF16 RTN & 0.837 & 0.033 & 0.837 & 0.016 \\
        BF16 + SR & 0.993 & 0.039 & 0.993 & 0.019 \\
        \midrule
        \multicolumn{5}{l}{\textit{Uniform precision}} \\
        \midrule
        FP16 RTN & 0.999 & 0.036 & 0.999 & 0.018 \\
        FP16 + SR & 0.441 & 0.056 & 0.441 & 0.028 \\
        \textbf{BF16 RTN} & $\mathbf{\approx 0}$ & \textbf{0.365} & $\mathbf{\approx 0}$ & \textbf{0.183} \\
        BF16 + SR & 0.072 & 0.108 & 0.072 & 0.054 \\
        \bottomrule
    \end{tabular}
    \caption{Results for the dimerisation model: monomer ($A$) and dimer ($D$) distributions. The conservation law confers numerical stability across most modes; uniform BF16 RTN shows moderate deviation rather than the catastrophic failure seen in other models. Bold entries indicate qualitative failure (distributional corruption or overflow).}
    \label{tab:dimer}
\end{table}

\begin{figure}[ht]
    \centering
    \includegraphics[width=\textwidth]{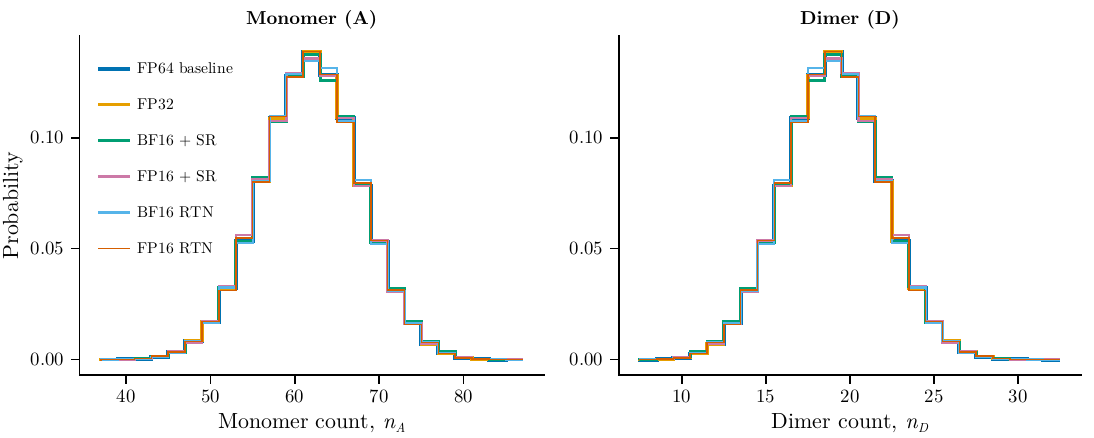}
    \caption{Steady-state distributions for monomer $A$ (left) and dimer $D$ (right) across all six mixed-precision modes. All configurations overlap closely with the FP64 baseline, consistent with the numerical stability conferred by the conservation law. Uniform-precision modes (not shown) are visually similar for most configurations, though uniform BF16 RTN shows moderate deviation; see Table~\ref{tab:dimer} for quantitative comparison.}
    \label{fig:dimer}
\end{figure}

Table~\ref{tab:dimer} and Figure~\ref{fig:dimer} showed that in mixed-precision mode every configuration passed both validation criteria comfortably: all KS $p$-values exceeded $0.5$ and all Wasserstein distances remained below $0.05$. FP32 was essentially exact, with $W_1 = 0.001$ for monomer $A$ and $W_1 = 0.0005$ for dimer $D$. Among the half-precision modes, FP16 RTN ($W_1 = 0.016$) outperformed FP16 + SR ($W_1 = 0.046$). This was the only model in our study where deterministic rounding matched or exceeded SR. BF16 RTN ($W_1 = 0.033$) likewise performed well, in contrast to its poor performance on the Telegraph and Schl\"{o}gl models.

A likely explanation for this robustness is the conservation law $n_A + 2n_D = 100$, which acts as an implicit error-correcting mechanism. The moderate population sizes and simple two-reaction structure may also contribute. In the SSA, state updates are integer-valued ($\pm 1$ or $\pm 2$), so the conservation constraint is enforced exactly at every step regardless of how propensities are rounded. This prevents the directional bias accumulation that causes RTN failure in the Telegraph and Schl\"{o}gl models: any rounding-induced distortion of a propensity may shift the relative probability of the next reaction, but the resulting state transition still satisfies the constraint, confining the system to the correct manifold.

Uniform-precision mode (Table~\ref{tab:dimer}) confirmed that the conservation law's protective effect extended even to all-reduced-precision operation. Uniform FP16+SR ($W_1 = 0.056$, KS $p = 0.44$) and uniform FP16 RTN ($W_1 = 0.036$, KS $p = 0.999$) both passed validation comfortably. Even uniform BF16 RTN, which failed catastrophically on the birth--death and Telegraph models, produced only moderate deviation here ($W_1 = 0.37$, KS $p \approx 0$); the conservation constraint prevents the runaway bias that destroys other models. The dimer is the only model in our study where uniform RTN remains viable, illustrating that structural constraints in the reaction network can provide inherent numerical stability that reduces the advantage of SR even at half precision.

\subsection{Repressilator}
\label{sec:repressilator}

The repressilator \cite{elowitz2000} is a synthetic three-gene oscillatory circuit in which three repressor proteins form a cyclic inhibition loop ($A \dashv B \dashv C \dashv A$), producing sustained oscillations. The stochastic model comprises six species (three mRNAs $M_i$ and three proteins $P_i$, $i \in \{A,B,C\}$) governed by four reactions per gene:
\begin{align*}
    \varnothing &\xrightarrow{f(P_j)} M_i, \quad M_i \xrightarrow{\delta_m} \varnothing, \quad M_i \xrightarrow{\beta} M_i + P_i, \quad P_i \xrightarrow{\delta_p} \varnothing,
\end{align*}
where $f(P_j) = \alpha_0 + \alpha/(1+P_j^h)$ is the Hill-regulated transcription rate and the repressor index cycles as $(j_A, j_B, j_C) = (C, A, B)$. We use $\alpha_0 = 1$, $\alpha = 216$, $\delta_m = 1$, $\beta = 5$, $\delta_p = 1$, and $h = 2$. The initial condition is asymmetric ($p_A(0) = 5$, all other species zero) to break the three-fold symmetry and seed oscillations. Under these parameters the system exhibits sustained stochastic oscillations with protein peaks ranging from roughly 50 to 200 molecules. The high event rate (${\sim}400$ reactions per time unit) makes uniform RTN modes particularly susceptible to time-step stagnation, as discussed in the birth--death uniform-mode analysis. The repressilator therefore tests a distinct failure mode from the previous models: long-time error accumulation in stiff oscillatory dynamics, where any persistent rounding bias is amplified over many oscillation cycles rather than simply shifting a steady-state mean. This amplitude range sits comfortably within FP16's representable range. Oscillators with much wider amplitude, spanning many orders of magnitude, would face the same dynamic-range constraints encountered with the original Schl\"{o}gl parameterisation; standard nondimensionalisation would restore FP16 viability there as well.

\begin{table}[ht]
    \centering
    \begin{tabular}{@{}lcccc@{}}
        \toprule
        Precision Mode & KS $p$-value & Wasserstein & Mean & Variance \\
        \midrule
        \multicolumn{5}{l}{\textit{Mixed precision}} \\
        \midrule
        FP64 (baseline) & 1.0 & 0 & 55.7 & 11944 \\
        FP32 & 0.573 & 4.76 & 51.0 & 10443 \\
        FP16 RTN & 0.828 & 3.15 & 54.2 & 9927 \\
        FP16 + SR & 0.913 & 4.16 & 53.0 & 10044 \\
        BF16 RTN & 0.859 & 4.26 & 52.5 & 10773 \\
        BF16 + SR & 0.263 & 5.57 & 50.4 & 10491 \\
        \midrule
        \multicolumn{5}{l}{\textit{Uniform precision}} \\
        \midrule
        \textbf{FP16 RTN} & $\mathbf{1.1 \times 10^{-5}}$ & \textbf{61.6} & \textbf{116.0} & \textbf{46831} \\
        FP16 + SR & 0.828 & 4.55 & 52.0 & 10697 \\
        \textbf{BF16 RTN} & $\mathbf{\approx 0}$ & \textbf{721.9} & \textbf{777.6} & \textbf{155042} \\
        \textbf{BF16 + SR} & $\mathbf{4.2 \times 10^{-4}}$ & \textbf{66.5} & \textbf{122.1} & \textbf{63510} \\
        \bottomrule
    \end{tabular}
    \caption{Results for the repressilator protein~A stationary statistics ($t_{\mathrm{end}} = 200$, $N = 10{,}000$ realisations); the stationary distribution is the time-invariant ensemble distribution of $p_A$ that emerges despite oscillating individual trajectories. By the three-fold symmetry of the circuit, proteins B and C have the same marginal distribution. In uniform mode, BF16+SR exhibits severe drift; RTN modes fail catastrophically. Bold entries indicate qualitative failure (distributional corruption or overflow).}
    \label{tab:repressilator}
\end{table}

\begin{figure}[ht]
    \centering
    \begin{subfigure}[t]{0.48\textwidth}
        \centering
        \includegraphics[width=\textwidth]{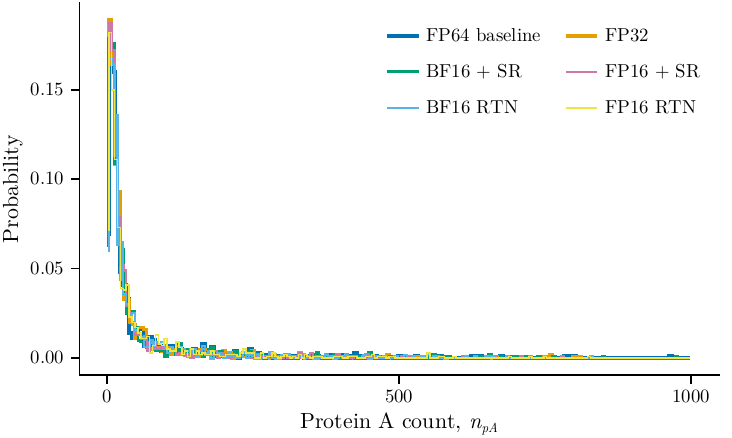}
        \caption{Mixed precision}
    \end{subfigure}
    \hfill
    \begin{subfigure}[t]{0.48\textwidth}
        \centering
        \includegraphics[width=\textwidth]{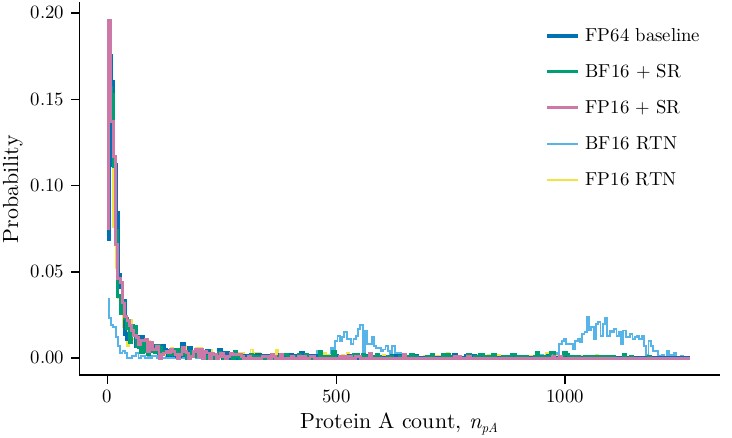}
        \caption{Uniform precision}
    \end{subfigure}
    \caption{Stationary protein~A distributions for the repressilator.
    (a)~Mixed-precision modes overlap closely with the FP64 baseline.
    (b)~Uniform-mode comparison: FP16+SR remains faithful, while BF16+SR ($W_1 = 66.5$) and RTN modes ($W_1 = 61.6$--$721.9$) fail catastrophically.}
    \label{fig:repressilator-dist}
\end{figure}

\begin{figure}[ht]
    \centering
    \includegraphics[width=\textwidth]{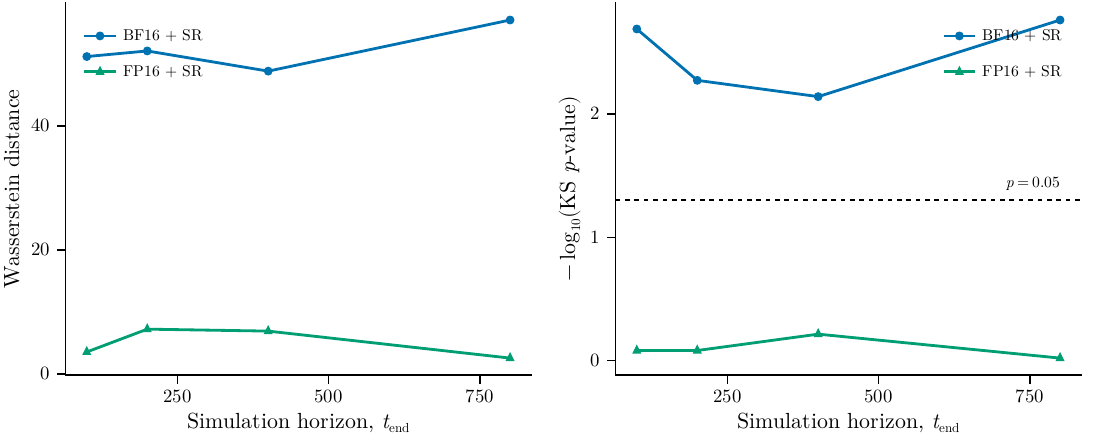}
    \caption{Repressilator horizon sweep under uniform precision ($t_{\mathrm{end}} = 100, 200, 400, 800$). Left: Wasserstein distance to FP64 baseline. Right: $-\log_{10}$ KS $p$-value, with the horizontal dashed line marking the $p = 0.05$ significance threshold. BF16+SR drift is systematic and horizon-independent ($W_1 \approx 49$--$57$; KS $p < 0.01$ at every horizon), while FP16+SR remains close to the baseline ($W_1 \approx 2.5$--$7.2$; KS $p > 0.6$).}
    \label{fig:repressilator-drift}
\end{figure}

In mixed-precision mode (Table~\ref{tab:repressilator}), all six configurations passed both validation criteria. KS $p$-values ranged from $0.263$ (BF16+SR) to $0.913$ (FP16+SR), and Wasserstein distances lay between $W_1 = 3.15$ (FP16 RTN) and $W_1 = 5.57$ (BF16+SR), corresponding to roughly 6--10\% of the baseline mean of 55.7 molecules. FP32 occupied the middle of this range ($W_1 = 4.76$, KS $p = 0.573$), indicating that even single-precision arithmetic introduced measurable distributional shift in this oscillatory system. As with the dimer model, FP16 RTN outperformed FP16+SR ($W_1 = 3.15$ vs.\ $4.16$), suggesting that the unbiased noise injected by SR can itself act as a perturbation source when propensity evaluation is confined to mixed precision. The moderate Wasserstein distances across all modes, larger than those seen in the birth-death and dimer models, are attributable to the inherent amplification of small perturbations by the oscillatory dynamics.

Uniform-precision mode revealed a qualitative failure boundary (Table~\ref{tab:repressilator} and Figure~\ref{fig:repressilator-dist}). Uniform BF16+SR failed decisively: the mean shifted from 55.7 to 122.1 (an increase of $119\%$), the variance inflated more than fivefold (from $11{,}944 $ to $63{,}510$), and the KS $p$-value dropped to $4.2 \times 10^{-4}$, below $p = 0.05$, with $W_1 = 66.5$. Uniform RTN modes were even worse: uniform BF16 RTN produced a mean of 777.6 ($W_1 = 721.9$, KS $p \approx 0$), reflecting runaway time stagnation that prevents the simulation from reaching stationarity. Uniform FP16 RTN ($W_1 = 61.6$, mean $= 116.0$) also failed, though less catastrophically. In contrast, uniform FP16+SR passed comfortably (KS $p = 0.828$, $W_1 = 4.55$, mean $= 52.0$), with variance comparable to the baseline ($10{,}697$). The mechanism underlying this divergence is mantissa resolution: BF16 carries a 7-bit mantissa while FP16 carries 10 bits. The Hill-function propensities $\alpha / (1 + p^h)$ with $p \in [0, 200]$ span several orders of magnitude, and BF16's coarser discretisation accumulates a systematic upward bias in protein counts over many oscillation cycles. SR removed the expected bias at each individual operation, but with only 7 mantissa bits the per-step rounding variance is large enough that error cancellation is incomplete over the long correlation times of the oscillator. This made the repressilator the only model in our study where SR was necessary but not sufficient for one format: BF16+SR fails while FP16+SR succeeds, showing that mantissa width, not just rounding mode, is a binding constraint for oscillatory kinetics.

Figure~\ref{fig:repressilator-drift} reinforced this conclusion with a horizon sweep over $t_{\mathrm{end}} \in \{100, 200, 400, 800\}$. Uniform BF16+SR maintained a Wasserstein distance of $49$--$57$ and KS $p < 0.01$ at every horizon, indicating that the drift was systematic rather than transient; it was present from the earliest time point and did not grow or shrink with longer simulation. Uniform FP16+SR, by contrast, remained close to the baseline at all horizons ($W_1 = 2.5$--$7.2$; KS $p > 0.6$). This horizon-independent character distinguishes the repressilator failure from the Telegraph model, where SR fully resolves precision-related artefacts: here, the combination of oscillatory amplification and BF16's limited mantissa creates a persistent bias that SR alone cannot eliminate. The practical implication is that FP16 (10-bit mantissa) is materially safer than BF16 (7~bits) for stiff oscillatory systems requiring long-time integration.

\subsection{Summary of numerical experiments}

Table~\ref{tab:summary} summarises Wasserstein distances across all models and precision modes.

\begin{table}[ht]
    \centering
    \small
    \begin{tabular}{@{}lccccccccl@{}}
        \toprule
        & \multicolumn{5}{c}{Mixed Precision} & \multicolumn{4}{c}{Uniform Precision} \\
        \cmidrule(lr){2-6} \cmidrule(lr){7-10}
        Model & FP32 & FP16 RTN & FP16+SR & BF16 RTN & BF16+SR & FP16 RTN & FP16+SR & BF16 RTN & BF16+SR \\
        \midrule
        Birth--Death & 0 & 0 & 0 & 0 & 0 & 0.22 & 0.03 & \textbf{3.75} & 0.12 \\
        Schl\"{o}gl & 1.41 & OVF$^{a}$ & OVF$^{a}$ & 24.1 & 19.7 & OVF$^{a}$ & OVF$^{a}$ & \textbf{35.9} & 22.9 \\
        \quad(rescaled) & 1.41 & 7.52 & 3.92 & 24.1 & 15.8 & 8.95 & 2.76 & \textbf{35.9} & 31.1 \\
        Telegraph & 0 & 0.001 & 0.001 & 0.007 & 0.007 & \textbf{0.098} & 0.011 & \textbf{0.438} & 0.016 \\
        \quad(bimodal) & 0 & 0.069 & 0.069 & 0.052 & 0.052 & \textbf{8.2} & 0.041 & \textbf{10.0} & 0.107 \\
        Dimer & 0.001 & 0.016 & 0.046 & 0.033 & 0.039 & 0.036 & 0.056 & \textbf{0.365} & 0.108 \\
        Repressilator & 4.76 & 3.15 & 4.16 & 4.26 & 5.57 & \textbf{61.6} & 4.55 & \textbf{721.9} & \textbf{66.5} \\
        \bottomrule
    \end{tabular}
    \par\smallskip\footnotesize $^{a}$FP16 overflow with original parameters ($A = 10^5$); viable after nondimensionalisation (rescaled row).
    \caption{Cross-model Wasserstein-1 distance summary (units: molecules). Bold entries indicate qualitative failure (distributional corruption or overflow).}
    \label{tab:summary}
\end{table}

The experiments across all five models yield four principal findings.
\begin{enumerate}
    \item \textbf{Mixed precision is broadly safe.} Using reduced-precision propensities with FP32 accumulators produced statistically valid results across all tested models. FP16 and FP32 modes were indistinguishable from FP64; BF16 modes showed moderate Wasserstein deviations on the Schl\"{o}gl model ($W_1 \approx 20$) but preserved the qualitative distributional structure.
    \item \textbf{SR is essential for uniform mode.} Uniform BF16 RTN fails on every model: birth--death (time stagnation, $W_1 = 3.75$), Telegraph (propensity swamping, $W_1 = 0.44$--$10.0$), Schl\"{o}gl (rounding bias feedback, $W_1 = 35.9$), dimer ($W_1 = 0.37$), and repressilator models (runaway divergence, $W_1 = 721.9$). Uniform FP16 RTN fails on the Telegraph, Schl\"{o}gl, and repressilator models but passes on the birth--death and dimer models. Uniform SR modes pass for all models except BF16+SR on the repressilator.
    \item \textbf{Model structure affects precision sensitivity:}
    \begin{itemize}
        \item Linear kinetics (birth--death): Mixed mode is trivially robust; uniform BF16 RTN fails via time stagnation.
        \item Conservation laws (dimer): Inherent numerical stability protects most uniform modes; even uniform BF16 RTN shows only moderate deviation.
        \item Oscillators (repressilator): Sensitive to long-time error accumulation; only uniform FP16+SR passes.
        \item High-order reactions (Schl\"{o}gl): FP16 overflow risk, recoverable via nondimensionalisation; rescaled uniform FP16+SR passes.
    \end{itemize}
    \item \textbf{SR is necessary but not always sufficient.} For long-time oscillatory simulations, BF16's limited mantissa (7 bits) causes drift even with SR. FP16 (10-bit mantissa) performs better in these cases.
\end{enumerate}

\subsection{Computational Performance}

We report central processing unit (CPU) benchmark results for all precision configurations across all five models, measuring median wall-clock time per trajectory, event throughput (SSA events per second), and memory usage. All benchmarks use mixed-precision mode (16-bit propensities with FP32 accumulators) and are run on a single CPU core with fixed ensemble sizes and simulation parameters matching the validation experiments above. Because the SSA is memory-bound rather than compute-bound on CPUs, and because current CPUs lack native 16-bit arithmetic units, substantial speedups are not expected in this setting. The primary performance opportunity lies in GPU and TPU implementations with native half-precision hardware, as discussed in Section~\ref{sec:discussion}.

\begin{table}[ht]
    \centering
    \small
    \begin{threeparttable}
    \begin{tabular}{@{}lccccc@{}}
        \toprule
        Model & FP32 & FP16 RTN & FP16+SR & BF16 RTN & BF16+SR \\
        \midrule
        Birth--Death   & 1.49$\times$ & 1.31$\times$ & 1.20$\times$ & 1.52$\times$ & 1.34$\times$ \\
        Schl\"{o}gl\textsuperscript{b} & 1.27$\times$ & 0.65$\times$ & 0.91$\times$ & 2.04$\times$ & 1.79$\times$ \\
        Telegraph      & 0.98$\times$ & 1.08$\times$ & 0.88$\times$ & 1.45$\times$ & 0.95$\times$ \\
        Dimer          & 1.08$\times$ & 1.13$\times$ & 1.08$\times$ & 1.11$\times$ & 1.08$\times$ \\
        Repressilator  & 1.08$\times$ & 0.91$\times$ & 0.76$\times$ & 0.77$\times$ & 0.84$\times$ \\
        \bottomrule
    \end{tabular}
    \begin{tablenotes}
        \footnotesize
        \item[a] FP64 baseline throughput (events/s): Birth--Death 621,320; Schl\"{o}gl 153,806; Telegraph 1,955,082; Dimer 1,398,319; Repressilator 476,167.
        \item[b] Schl\"{o}gl timings dominated by variable garbage-collection/allocation overhead; see text for details.
    \end{tablenotes}
    \end{threeparttable}
    \caption{Throughput speedup relative to FP64 baseline for each model under mixed-precision mode.\textsuperscript{a} Schl\"{o}gl numbers reflect anomalous memory-allocation behaviour.}
    \label{tab:benchmarks}
\end{table}

As shown in Table~\ref{tab:benchmarks}, throughput improvements from reduced precision were modest on CPU hardware. The birth--death model showed the largest reliable gains, with BF16 RTN achieving 1.52$\times$ improvements and FP32 reaching 1.49$\times$, consistent with its simple propensity structure and high event rate. The Telegraph model's BF16 RTN mode achieved a notable 1.45$\times$ speedup, while the dimer showed uniformly modest gains (1.08--1.13$\times$). The repressilator showed negligible or negative speedups across all 16-bit modes (0.76--0.91$\times$), with only FP32 achieving a modest gain (1.08$\times$). The Schl\"{o}gl model exhibited anomalous benchmark behaviour: memory footprints varied dramatically across precision modes (from 40\,MB for BF16 RTN to 2.9\,GB for FP16 RTN), suggesting that garbage-collection and memory-allocation artefacts dominate the timing rather than arithmetic throughput. We report these numbers in Table~\ref{tab:benchmarks} for completeness but caution against drawing performance conclusions from them.

Two factors explain the modest CPU speedups. First, current CPUs lack native FP16/BF16 arithmetic, so each 16-bit operation requires promotion to FP32, computation, and demotion back to 16 bits, introducing overhead that can outweigh memory savings. SR modes incur a further penalty because SR is emulated in software, requiring additional random number generation at each rounding step. Second, the SSA is memory-bound rather than compute-bound: each step performs modest arithmetic while accessing state vectors, propensity arrays, and stoichiometry structures with irregular memory access patterns. Shrinking data representations provides limited benefit when the bottleneck is cache misses and memory latency rather than arithmetic logic unit (ALU) throughput. These overheads are most pronounced for complex models like the repressilator, where the larger state space amplifies both the emulation cost and the memory-access bottleneck. The true performance opportunity lies in GPU and TPU implementations, where native 16-bit ALUs and halved memory bandwidth requirements can deliver substantially larger gains, particularly for naively parallel ensemble simulations. Importantly, mixed-precision SSA is orthogonal to algorithmic acceleration (tau-leaping, next-reaction methods), meaning the two approaches can be combined.

\section{Discussion}
\label{sec:discussion}

The results across five canonical biochemical kinetics models converge on a clear conclusion: mixed-precision SSA is a viable strategy for common models in mathematical biology, with SR playing a critical enabling role when precision is pushed to its limits.

Our experiments suggest a straightforward partitioning for practitioners. Propensity calculations, i.e.\ multiplications of rate constants by population counts, can use 16-bit arithmetic (\texttt{fp16} or \texttt{bf16}) with minimal impact on ensemble statistics for most systems considered here; some edge cases are discussed below. Accumulation operations (summing propensities, advancing the simulation clock) should remain in at least \texttt{fp32}. This mixed-precision strategy produced results that closely matched FP64 across all five validated models, including systems with nonlinear propensities (dimer), rare events (Telegraph), bistability (Schl\"{o}gl), and stiff oscillatory dynamics (repressilator). BF16 modes on the Schl\"{o}gl model showed moderate distributional shifts ($W_1 \approx 20$), but preserved the bimodal structure and did not corrupt the qualitative dynamics.

The choice between FP16 and BF16 depends on system characteristics. FP16, with its 10-bit mantissa, offers better long-time fidelity: the repressilator in uniform FP16+SR mode achieved a Wasserstein distance of 4.55, compared to 66.5 for BF16+SR. However, BF16's wider dynamic range (8-bit exponent) makes it the only viable 16-bit option when propensities exceed FP16's maximum of 65,504, as occurs in the Schl\"{o}gl model's cubic propensities. When propensities can be rescaled into FP16's representable range, FP16+SR should be preferred for its precision advantage.

The decision between mixed and uniform precision modes is equally important. Mixed precision (16-bit propensities with FP32 accumulators) was safe across all tested models and should be the default recommendation. Uniform mode requires SR and is validated only for systems without stiff long-time dynamics. Uniform FP16+SR is appropriate for linear kinetics (birth--death), systems with conservation laws (dimer), and moderate switching dynamics (Telegraph). Notably, even the rescaled Schl\"{o}gl model passes in uniform FP16+SR ($W_1 = 2.76$, KS $p = 0.81$), confirming that rescaling unlocks uniform mode for systems that would otherwise overflow. For stiff oscillators and long-horizon simulations, mixed mode remains the only reliable option. In short, models with simple or self-correcting propensity structures tolerate uniform mode, while models with complex multi-species dynamics or long-time integration require mixed-precision accumulation.

Across the five models studied, model structure appears to be a primary determinant of precision sensitivity. The five models span a spectrum: linear kinetics (birth--death) are trivially robust, conservation laws (dimer) provide implicit error correction, and switching dynamics (Telegraph) expose the critical role of SR when small propensities risk underflow. Two broader lessons emerge from the results. First, SR is necessary for uniform mode because it prevents the systematic rounding-to-zero of small propensities that causes RTN to fail (Section~\ref{sec:experiments}). Second, SR is not always sufficient: the repressilator under uniform BF16+SR demonstrates that mantissa width, not just rounding mode, determines long-time fidelity in oscillatory systems. The Schl\"{o}gl model presents a complementary limit: hard dynamic-range overflow that no rounding strategy can overcome, but that standard nondimensionalisation resolves. These patterns are drawn from a deliberately diverse but small set of canonical models; broader systematic studies (e.g., across larger reaction networks, alternative parameter regimes, and different propensity functional forms) would be needed to validate them as general rules.

A natural question is whether SR-induced rounding noise could interfere with the intrinsic molecular noise that SSA is designed to capture. In every model where SR modes pass the KS test, the Wasserstein distance is comparable to or smaller than the statistical sampling error of the ensemble itself. The Telegraph model is instructive: the SR modes (Wasserstein $\sim$0.01) show perturbations far smaller than the intrinsic standard deviation of the steady-state distribution ($\sigma \approx 1.5$). SR noise is effectively invisible against the molecular noise background, consistent with the analogy to weather and climate modelling where rounding noise is dominated by parametric and observational uncertainty \cite{palmer2015,chantry2019,kimpson2024}.

As discussed in Section~\ref{sec:experiments}, the modest CPU speedups (0.8--1.5$\times$) reflect two factors: the lack of native 16-bit arithmetic on current CPUs, which introduces promotion/demotion overhead, and the memory-bound nature of SSA, which limits the benefit of smaller data types. Mainstream NVIDIA GPUs do not currently support SR natively, though a growing number of AI accelerators do: the Graphcore IPU~\cite{graphcore2020}, Intel Gaudi~\cite{intelgaudi2024}, and AWS Trainium all implement hardware-accelerated SR for low-precision arithmetic. Efficient software algorithms for SR on conventional architectures have also been developed~\cite{fasi2021}. On hardware with native SR support, the emulation overhead would vanish. The real performance opportunity lies in GPU and TPU implementations, where native 16-bit ALUs, halved memory bandwidth requirements, and massive parallelism over ensemble members could deliver substantially larger gains. Mixed-precision SSA is also orthogonal to algorithmic acceleration (tau-leaping, next-reaction methods), meaning the two approaches can be composed.

Several limitations should be noted. First, our evaluation is CPU-only; GPU benchmarks with native \texttt{fp16} hardware are needed to quantify the full performance benefit. Second, ensemble sizes for the repressilator (1,000 trajectories) are smaller than for other models, reducing statistical power. Third, we use ``every-op'' SR via \texttt{StochasticRounding.jl}, which is the most conservative strategy; selective SR (applied only at accumulation points) might achieve similar accuracy with lower overhead but has not been tested here. Fourth, we have not explored the interaction between reduced precision and approximate SSA variants such as tau-leaping, where algorithmic approximation error and rounding error may interact in non-trivial ways. Finally, the models studied here are small (2--12 reactions, 1--6 species); larger reaction networks representative of signalling pathways or whole-cell models may introduce new precision challenges, such as wider propensity dynamic ranges or longer correlation times, that are not captured by these benchmarks.

While this study focuses on the well-stirred SSA, we expect the principles identified here to extend to spatial stochastic frameworks, though this remains to be validated. The RDME \cite{engblom2009} couples SSA-like kinetics within lattice subvolumes to stochastic diffusive transport between neighbouring compartments. The local reaction step is arithmetically identical to the SSA studied here, so mixed-precision propensity evaluation should carry over to each subvolume, with accumulation remaining in FP32. The diffusive transport step introduces additional precision requirements (computing diffusion propensities across many subvolumes, maintaining spatial conservation) that would need separate investigation. The larger state vectors in RDME simulations, which scale with the number of spatial compartments, would amplify the memory savings from reduced precision.

Agent-based models in epidemiology and tissue modelling \cite{an2009} share the discrete stochastic character that makes reduced precision plausible: transition rates between agent states are analogous to propensities, and similar precision-map principles may apply. Hybrid stochastic-deterministic models present a natural decomposition: the SSA component can use the mixed-precision strategy validated here, while the ODE component may benefit from reduced precision independently. In all these frameworks, intrinsic stochastic noise provides a natural tolerance for rounding error, just as it does in the well-stirred case.

Deterministic reaction-diffusion PDE models \cite{hundsdorfer2003}, widely used in mathematical biology, lack intrinsic noise to mask rounding error, but they are memory-bound on fine spatial grids and could benefit from halved data traffic. Precedent exists in computational fluid dynamics, where mixed-precision solvers have demonstrated substantial speedups without sacrificing solution quality \cite{openfoamFP2023}. For deterministic solvers, rounding error bounds would need to be established through traditional numerical analysis rather than the statistical testing approach used here. The precision maps and SR guidelines established in this work for SSA provide a starting point for these extensions.

\section{Conclusion}
\label{sec:conclusion}

We have presented the first systematic investigation of reduced-precision arithmetic for stochastic simulation in mathematical biology. Our results demonstrate that mixed-precision SSA, computing propensities in 16-bit formats while maintaining accumulators in 32-bit precision, preserves statistical fidelity across a range of canonical biochemical kinetics models, from simple linear kinetics to stiff oscillatory circuits.

Four principal findings emerge from this study. First, mixed precision is safe across all tested models: reduced-precision propensities with FP32 accumulators produce ensemble statistics indistinguishable from the FP64 reference. Second, SR is essential for uniform reduced-precision operation, as demonstrated most clearly by the Telegraph model, where SR maintains distributional fidelity (KS $p > 0.94$) while round-to-nearest fails catastrophically (KS $p \approx 0$). Third, model structure determines precision sensitivity: linear kinetics are trivially robust, conservation laws provide implicit error correction, and stiff oscillators are vulnerable to long-time drift. Fourth, SR is necessary but not always sufficient. For long oscillatory simulations, mantissa precision also matters, and BF16's 7-bit mantissa is insufficient even with SR.

These findings establish reduced-precision arithmetic as a viable, orthogonal axis for accelerating stochastic simulation, complementary to existing algorithmic strategies. Future work should target GPU implementations with native 16-bit hardware, where the combination of halved memory traffic and massive ensemble parallelism promises substantially larger speedups. Other natural extensions include combining mixed precision with tau-leaping, developing rate-rescaling strategies for models with large propensities, and scaling to larger reaction networks representative of whole-cell models. Beyond well-stirred systems, extending these ideas to spatial stochastic frameworks such as the RDME, agent-based models, and memory-bound deterministic PDE solvers represents a natural next step toward reduced-precision simulation across mathematical biology.

\section*{Code and Data Availability}

All code, simulation scripts, and raw results are available at \url{https://github.com/tomkimpson/LowPrecBio}. 

\bibliographystyle{unsrtnat}
\bibliography{references}

\end{document}